\documentclass[prd,twocolumn,reprint,preprintnumbers,nofootinbib]{revtex4-1}

\input{header}

\begin{document}
\title{Forward Neutrino Fluxes at the LHC}

\author{Felix Kling}
\email{felixk@slac.stanford.edu}
\affiliation{Theory Group, SLAC National Accelerator Laboratory, Menlo Park, California 94025}

\author{Laurence J. Nevay}
\email{laurie.nevay@rhul.ac.uk}
\affiliation{Royal Holloway, University of London, Egham TW20 0EX, United Kingdom}

\begin{abstract}
 With the upcoming Run~3 of the LHC, the FASER$\nu$ and SND@LHC detectors will start a new era of neutrino physics using the far-forward high-energy neutrino beam produced in collisions at ATLAS. This emerging LHC neutrino physics program requires reliable estimates of the LHC’s forward neutrino fluxes and their uncertainties. In this paper we provide a new fast-neutrino flux simulation, implemented as a RIVET module, to address this issue. We present the expected energy distributions going through the FASER$\nu$ and SND@LHC detectors based on various commonly used event generators, analyze the origin of those neutrinos, and present the expected neutrino event rates.   
\end{abstract}

\maketitle

\section{Introduction}
\label{sec:intro}

As the highest energy particle accelerator ever built, the Large Hadron Collider (LHC) is also the source of the most energetic neutrinos created in a controlled laboratory environment. Proton-proton collisions at the LHC typically lead to a large number of hadrons produced along the beam direction, which can inherit a significant fraction of the protons’ energy. The decays of these hadrons then lead to an intense and strongly collimated beam of highly energetic neutrinos of all three flavors in the far-forward direction.   

Although studies on the possibility of detecting and probing neutrinos at the LHC and associated neutrino fluxes reach back to 1984~\cite{DeRujula:1984pg, Vannucci:253670, DeRujula:1992sn, Park:2011gh, Buontempo:2018gta, Beni:2019gxv, XSEN:2019bel, Beni:2020yfy}, no LHC neutrino has been detected until very recently. This situation changed when the FASER collaboration reported the observation of the first neutrino interaction candidates at the LHC~\cite{Abreu:2021hol}. This situation will further improve during the third run of the LHC with upcoming FASER$\nu$~\cite{Abreu:2019yak, Abreu:2020ddv} and SND@LHC detectors~\cite{Ahdida:2020evc, Ahdida:2750060}. Placed directly in the LHC's forward neutrino beam, both experiments are expected to detect thousands of neutrino interactions at TeV energies. This will open a new window to study neutrino interactions at high energies and therefore extend the LHC’s physics program in a new direction.

This emerging LHC neutrino physics program requires reliable estimates of the LHC's forward neutrino fluxes and their uncertainties. These estimates are typically based on established Monte Carlo (MC) event generators to simulate the production of hadrons in proton-proton interactions. For these hadrons we then have to simulate the propagation through the LHC's beam pipe and magnetic fields as well as their decay into neutrinos. This can be done using dedicated simulation tools such as BDSIM~\cite{Nevay:2018zhp} or FLUKA~\cite{Ferrari:2005zk}. However, these simulations tend to be rather computationally expensive, time consuming, and often require special expertise or code access that is not available to the broad community. This makes the simulation of neutrino fluxes with different generators, as, for example, needed to obtain flux uncertainties or for phenomenological studies, difficult to impossible. In this study we address this issue and present an alternative fast neutrino flux simulation implemented as a RIVET~\cite{Buckley:2010ar, Bierlich:2019rhm} module\footnote{The module and fluxes presented in this paper are available at \href{https://github.com/KlingFelix/FastNeutrinoFluxSimulation}{https://github.com/KlingFelix/FastNeutrinoFluxSimulation}}.

The rest of this paper is organized as follows. In \cref{sec:simulation}, we provide a description of our fast neutrino flux simulation module. In \cref{sec:fluxes} we present the neutrino energy and rapidity spectra obtained from different event generators. In \cref{sec:interactions} we present the expected neutrino interactions rates in FASER$\nu$ and SND@LHC and discuss the effect of the beam crossing angle. We conclude in \cref{sec:summary}. 

\section{Fast Neutrino Flux Simulation}
\label{sec:simulation} 

\begin{figure*}[th!]
\includegraphics[width=1.0\textwidth]{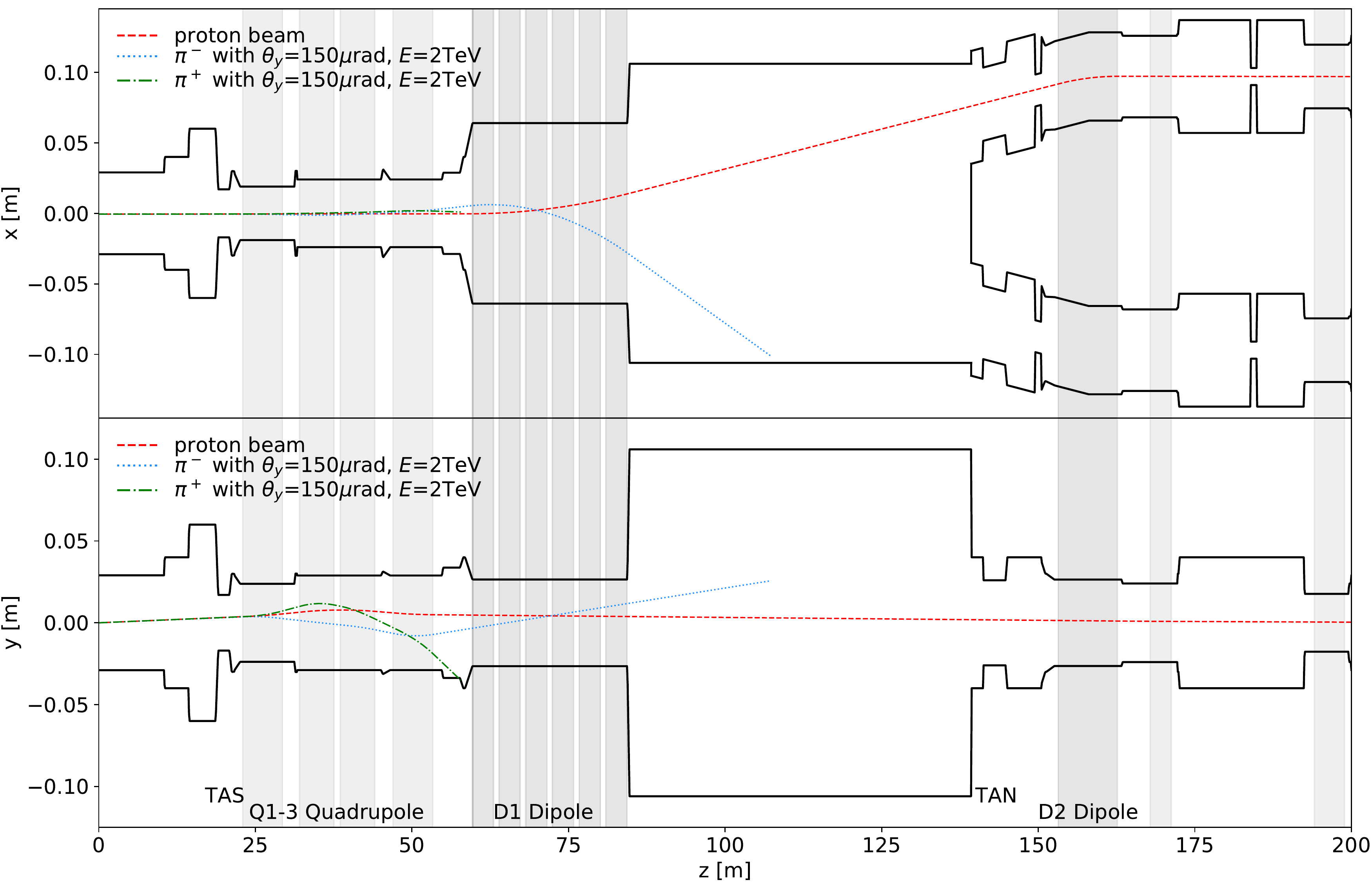}
\caption{\textbf{Beam Pipe Geometry:} The boundaries of the LHC's beam pipe in the horizontal (top) and vertical (bottom) planes are shown as a black line. The shaded areas correspond to the quadrupole (light gray) and dipole (dark gray) magnets. The red dashed line shows the trajectory of the proton beam with initial half beam crossing angle of $150~\murad$ vertically upwards. The dotted blue and dot-dashed green curve show the trajectories of two oppositely charged 2~TeV energy pions with the same initial orientations.}
\label{fig:geometry}
\end{figure*}

Neutrinos going through the far-forward LHC neutrino experiments mainly originate from the weak decay of the lightest mesons and baryons of a given flavor. At the LHC, this includes the decay of pions, kaons, hyperons, D-mesons and charmed baryons. When simulating the forward LHC neutrino beam, we can distinguish three contributions based on the production location: i) a \textit{prompt} neutrino flux component from charm hadron decay occurring essentially at the collision point, ii) a \textit{displaced} neutrino flux component from the decay of light hadrons in the LHC vacuum beam pipe before they collide with material, and iii) a \textit{secondary} neutrino flux component from downstream hadronic showers resulting from collisions of primary hadrons with the LHC infrastructure. 

A full simulation of the LHC neutrino flux, taking into account all three components, requires both a geometrical description of the LHC infrastructure downstream of the collision point and the modeling of particle propagation through it. This is available in dedicated particle transport and interaction codes, such as FLUKA and BDSIM (which is based on \texttt{Geant4}). However, using these tools for LHC neutrino flux studies faces two challenges: i) Running a full simulation with sufficient statistics takes typically between 1000 to 10000 CPU-hours on a computer cluster, meaning that it is both computationally expensive and time consuming. ii) Using the tools requires some expertise to run them as well access to the code and geometrical model. This practically restricts running the simulations to the groups maintaining the code. Both of these challenges make it difficult for the broader community to use them for phenomenological studies and applications, for example when the neutrino flux has to be obtained for many different generators or generator configurations. 

The long running time of a full simulation is mainly associated with downstream hadronic showers, which lead to a large number of low-energetic particles whose propagation through material needs to be simulated. However, given the typically lower energy and large spread of hadrons in later stages of the shower, as well as their small probability to decay inside a dense medium, the resulting secondary neutrino flux component is expected to be subdominant, at least for higher energies. To quantify this effect, we have performed a full simulation using BDSIM interfaced with \texttt{Sibyll~2.3d}~\cite{Riehn:2019jet} to estimate the fraction of neutrinos from this secondary component\footnote{More details on a forward LHC neutrino flux simulation using BDSIM will be reported in a separate study~\cite{FASER-BDSIM}.}. We found that only about $0.4~\%$ ($1.5~\%$, $2.0~\%$, $4~\%$) of muon neutrinos with energy $E>1~\tev$ ($300~\gev$, $100~\gev$, $30~\gev$) passing through a $40~\cm \times 40~\cm$ cross sectional area at a location $z=480~\m$ downstream from the ATLAS IP originate from decays in a medium. This means that neutrinos, especially at the higher energies of interest, are primarily produced in the vacuum of the LHC beam pipe, while neutrinos from  downstream secondary interactions only contribute a small subdominant fraction to the neutrino flux.

Motivated by this finding, this study presents a \textit{fast neutrino flux simulation} that focuses on the dominant prompt and displaced neutrino flux components. This simplified approach only tracks particles inside the LHC's vacuum beam, but does not simulate their propagation through material, leading to a significant reduction in computation time. In the following, we present the geometrical model in \cref{sec:fnfs-model}; the tracking of particles through the LHC's magnetic fields in \cref{sec:fnfs-tracking}; the modeling of particle decays in \cref{sec:fnfs-decays}; a parametric ansatz to simulate parts of the sub-leading secondary neutrino flux in \cref{sec:fnfs-interactions}; and the implementation as RIVET module in \cref{sec:fnfs-rivet}. In \cref{sec:fnfs-validation} we then compare the results of the fast neutrino flux simulation to a full simulation performed using BDSIM. 

\subsection{Geometrical Model}
\label{sec:fnfs-model}

The geometrical model used in this study is based on the one implemented in BDSIM~\cite{Nevay:2018zhp}. BDSIM is a program that creates Geant4 models of particle accelerators using a library of parameterized accelerator geometry. It provides accurate magnetic fields and particle tracking for all particles through an accelerator for a given magnetic configuration (``optics''). A highly detailed model of LHC from ATLAS (`IP1') to approximately $500~\m$ downstream and this will be detailed in a future publication. The shape of the vacuum pipe aperture has been extracted from this simulation and implemented in the described fast simulation as illustrated in \cref{fig:geometry}. The upper and lower panels show a cross sectional view of the beam pipe geometry in the horizontal and vertical plane, respectively, as indicated by the solid black contour. Additionally, we show the quadrupole and dipole magnet regions as light and dark gray shaded areas, respectively.

Located $19.1~\m$ downstream from the ATLAS interaction point (IP) is the TAS front quadrupole absorber. It is a 1.8 m-long copper block with a $3.4~\cm$ aperture opening for the beam which absorbs hadrons with angles $\theta \gtrsim 0.9~\mrad$ with respect to the beam axis. The TAS is followed by a series of quadrupole magnets, the so called inner triplet, and the D1 magnet. The D1 dipole separates the two proton beams and also deflects most charged particles such that they collide with the beam pipe. Placed at $140~\m$ downstream from the IP is the TAN, which will absorb the forward going neutral particles. At this location, the beam pipe splits into separate pipes for the two beams. Further downstream, at $z=153~\m$, is the outer beam separation dipole magnet D2, which re-aligns the proton beams to be parallel. Placed at about $z = 150~\m$, $184~\m$ and $218~\m$ are three collimators of $1~\m$ length designed to absorb any tertiary proton beam halo incoming to the experiment and also absorb any slightly deflected protons or other high energy physics debris to protect the accelerator.

Within the simulation, particles hitting this boundary of the beam pipe are assumed to be absorbed quickly and are no longer tracked. This intuitively applies to regions where dense metal objects, like the TAS, the TAN, the magnets and the collimators, surround the beam pipe. The situation is less clear when the vacuum is only surrounded by the about $5~\mm$ thick beam pipe. This applies to the region between the D1 magnets and the TAN, which is mainly hit by energetic charged particles that were deflected by the D1 magnets. However, those particles typical hit the pipe at a few $\mrad$ angle, which is sufficiently small such that particles would need to travel through a few meters of material where they will likely interact. In addition, a few $\mrad$ angle is already quite large, such that neutrinos produced from their decay are typically not relevant for the far-forward neutrino fluxes. Indeed, we found that charged particle decays into neutrinos occurring after the dipole magnets only provide a small contribution to the neutrino flux.  

\subsection{Tracking through Magnetic Fields}
\label{sec:fnfs-tracking}

While charmed hadrons decay approximately promptly, light flavored hadrons are long-lived and decay downstream from the interaction point. This requires us to model their deflection by the magnetic fields to obtain their trajectories. Here we implement the \texttt{BDSIM Quadrupole} and \texttt{BDSIM Dipole Rodrigues} first order matrix tracking algorithms that are described in the BDSIM documentation~\cite{Nevay:2018zhp, BDSIM}; see also Ref.~\cite{Wiedemann:2015fja} for an pedagogical overview. 

Also shown in \cref{fig:geometry} is the trajectory for the $6.5~\tev$ proton beam with the nominal LHC Run~2 beam half-crossing angle $150~\murad$ vertically upwards, as well as two trajectories for oppositely charged 2~TeV energy pions. We have validated these and other trajectories against the full BDSIM prediction and found great agreement of $\mathcal{O}(10~\mu\m)$ and better. We can see that even charged particles with large energies of 2~TeV are deflected significantly by the inner triplet quadrupole magnets, such that further downstream decays of these charged particles are not expected to contribute significantly to the far-forward neutrino flux.

\subsection{Decays into Neutrinos}
\label{sec:fnfs-decays}

In the next step, we decay the hadrons to obtain the neutrino flux. Unfortunately, the hadron decay branching fractions differ between different event generators. Most notably, the decay tables for several dedicated simulators, including \texttt{Sibyll~2.3d} and \texttt{DPMJET~III.2017.1}, are incomplete and do not contain the decay channels $\Lambda \to p e \nu$ or $D_s \to \tau \nu_\tau$. To avoid this problem and allow for a fair comparison of different hadronic interaction models, the decays of hadrons into neutrinos is simulated within the module. As a basis, we use the decay branching fractions and the energy distributions of neutrinos in the hadron rest frame, as obtained by \texttt{Pythia~8}~\cite{Sjostrand:2014zea}. 

For light long-lived hadrons, we sample the decays of hadrons into neutrinos roughly every meter along their trajectory inside the beam pipe, taking into account the corresponding probabilities to decay in flight. Heavy hadrons are decayed promptly at the IP, where each hadron decay is simulated a thousand times to increase the sampling statistics in the far-forward direction. Both light and heavy hadrons are first decayed in their rest frame, according to the decay branching fractions and energy distributions obtained with \texttt{Pythia~8}, and the resulting neutrino is then boosted into the laboratory frame.

\begin{figure*}[th!]
\includegraphics[width=1.0\textwidth]{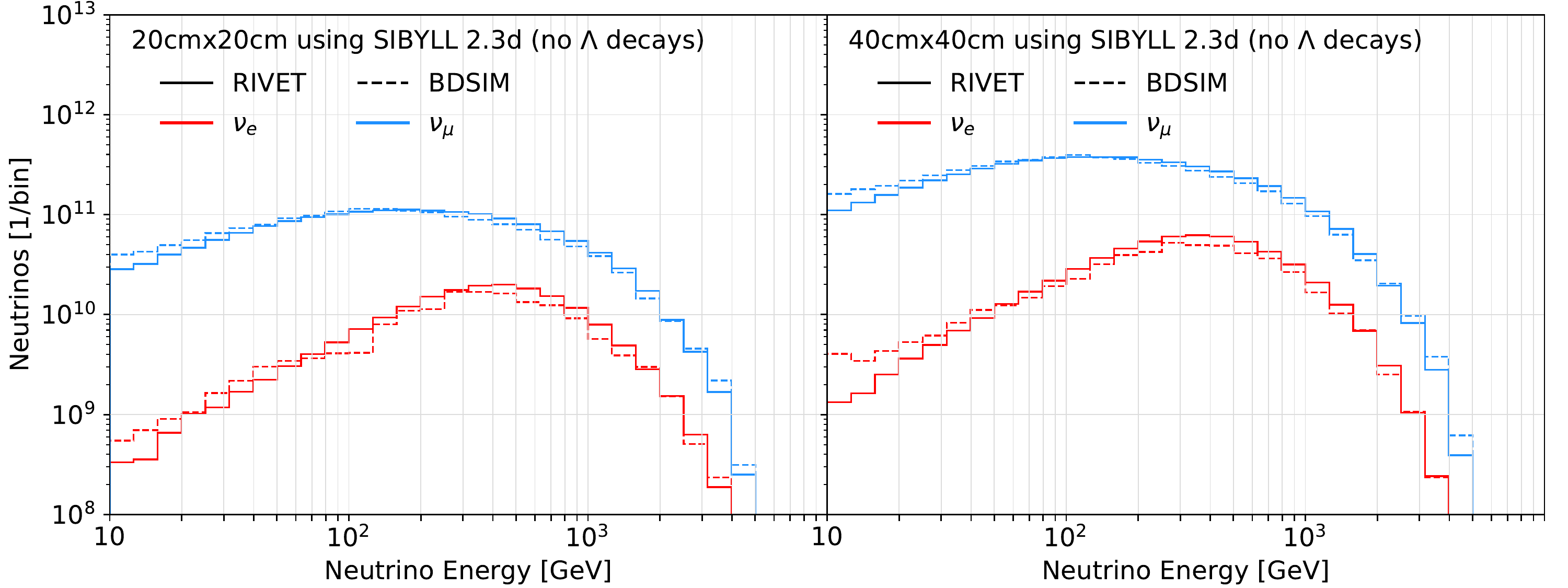}
\caption{\textbf{Validation:} Comparison of the neutrino flux obtained using the fast neutrino flux simulation implemented as a RIVET module (solid) and a full detector simulation using BDSIM (dashed) for LHC Run~3 with an integrated luminosity of $150~\ifb$. We show the energy spectrum of the neutrino passing through a  $20~\cm \times 20~\cm$ (left) and $40~\cm \times 40~\cm$ (right) cross sectional area that is centered around the beam collision axis and located at $z=480~\m$. Electron and muon neutrinos are shown in red and blue, respectively}
\label{fig:validation}
\end{figure*}

\subsection{Neutrinos from Interactions}
\label{sec:fnfs-interactions}

In addition to hadron decays inside the LHC's beam pipe, neutrinos can also be produced inside hadronic showers resulting from collisions of primary hadrons with the LHC infrastructure. Using the full BDSIM simulation, we have already seen that the corresponding secondary contribution to the far-forward neutrino flux is strongly suppressed. This is due to the small probability of secondary hadrons to decay in medium before interacting again as well as their typically broad angular spread.

Nevertheless, to illustrate the feasibility of including this component in a fast simulation, we include a simplified modeling of the secondary neutrino flux component. For this, we simulated the interactions of neutrons of different energies with a thick copper target in \texttt{Geant4}~\cite{Agostinelli:2002hh} using the \texttt{FTFP\_BERT} Geant4 reference physics list and recorded the resulting neutrino flux in the form of a two-dimensional energy-angle distribution. Here, the neutron beam and copper target were chosen to reproduce the setup of forward neutron collisions with the TAS and TAN, which provide the largest contribution to the secondary neutrino flux. In the fast simulation when a hadron collides with the LHC's beam pipe boundaries, we then sample the resulting neutrinos following the obtained distributions. This approach automatically takes into account neutrinos produced in the decay of particles from all stages of the hadronic shower. 

While the approach outlined above is targeting neutrinos produced inside dense materials, we note that there are other components of the secondary neutrino flux that are not yet included. This includes, for example, neutrinos from hadrons produced inside a material that enter either the beam pipe vacuum or the air where they have a much larger probability to decay. A dedicated study to understand the nature and importance of different secondary neutrino flux components using a full BDSIM simulation is underway~\cite{FASER-BDSIM}. These results can be used to improve the present fast neutrino flux simulation. 

\subsection{Implementation in RIVET}
\label{sec:fnfs-rivet}

All parts of the outlined simulation are implemented as a RIVET module. RIVET is a toolkit for \textit{Robust Independent Validation of Experiment and Theory}, which processes the output of event generators to obtain histograms. It contains a library of analysis modules that reproduce experimental analyses and therefore allow comparison of theoretical calculations for final state distributions to measurements. In this work, we have created a new module that predicts the neutrino flux that can be observed at the FASER$\nu$ and SND@LHC experiments, consisting of three sub-modules for the prompt, displaced and secondary neutrino flux components, respectively. 

This module i) reads the forward hadron fluxes from \texttt{HepMC}~\cite{Dobbs:2001ck} files produced by the MC generator, ii) propagates the long-lived hadrons through the LHC beam pipe and magnets iii) obtains the neutrinos from decays of hadrons at multiple locations along their trajectory or interactions of hadrons with the beam pipe material, and iv) stores the resulting neutrino fluxes going through the forward LHC neutrino experiments FASER$\nu$ and SND@LHC as histograms in the \texttt{yoda} file format~\cite{Buckley:2010ar}. The run time of the RIVET module is comparable to that of event generation, meaning that neutrinos from thousands of collisions can be simulated within minutes on a normal computer. This provides a significant improvement in computation time compared to the BDSIM or FLUKA simulations. 

\subsection{Validation}
\label{sec:fnfs-validation}

To validate the predictions of our fast neutrino flux simulation, we performed a comparison to a full simulation using BDSIM. This is shown in \cref{fig:validation}, where we plot the energy spectra of neutrinos passing through a smaller $20~\cm \times 20~\cm$ (left) and larger $40~\cm \times 40~\cm$ (right) cross sectional area that are located 480~m downstream from the ATLAS interaction point and centered around the beam collision axis. The predictions of the fast neutrino flux simulation and full simulation are shown as solid and dashed histograms, respectively.  

To allow a meaningful comparison, we use the same setup for both simulations. In particular, we consider 13~TeV LHC with a beam half-crossing angle of $150~\murad$ vertically upwards, use same geometrical description for the beam pipe geometry and magnets, and simulate the primary interactions with \texttt{Sibyll~2.3d}. However, we note that there are small remaining differences, for example due to the modeling of particle decays. An important example is the decay $\Lambda \to p e \nu$, which is not implemented in \texttt{Sibyll} and as a result also not included in the full simulation. We therefore exclude neutrinos from hyperon decays in the comparison. 

The predictions of the full and fast simulations agree over the considered energy range, for both electron and muon neutrinos. In particular, as we will see in the next section, the differences between the full and fast simulations are significantly smaller than the differences between different generators. Some discrepancy arises at lower energies $E \!\sim\! 10~\gev$, which becomes more pronounced when going away from beam collision axis. This suggests that additional contributions to the secondary neutrino flux component, which have not yet been included in the fast simulation, might become relevant in this regime.

A comparison with the results of a full FLUKA simulation, which have been presented in Refs.~\cite{XSEN:2019bel, Beni:2020yfy, Ahdida:2020evc, Ahdida:2750060}, is presented in \cref{sec:fluka}.

\section{Forward Neutrino Fluxes}
\label{sec:fluxes} 

\subsection{Experimental Setup}
\label{sec:experiment} 

In the upcoming Run~3 of the LHC, which is scheduled from 2022 to 2024, two new LHC experiments, FASER$\nu$ and SND@LHC, will start their operation and probe neutrinos at the LHC for the first time. 
 
The FASER experiment has been originally proposed to search for light long-lived particles at the LHC~\cite{Feng:2017vli, Feng:2017uoz, Feng:2018noy, Kling:2018wct, Berlin:2018jbm, Ariga:2018zuc, Ariga:2018uku, Ariga:2018pin, Ariga:2019ufm, Kling:2020mch}. Placed at its front is a dedicated neutrino detector, called FASER$\nu$, which consists of emulsion films interleaved with tungsten plates of total mass $1.2$~tons~\cite{Abreu:2019yak, Abreu:2020ddv}. This setup allows measurement of the neutrino energy and can identify the neutrino flavor based on the signature in the emulsion detector, and distinguish muon neutrino and anti-neutrinos interactions in combination with the downstream FASER spectrometer. The experiment is located about $480~\m$ downstream from the ATLAS IP in the previously unused side tunnel TI12. At this location, a trench was dug, allowing the whole detector to be centered on the beam collision axis. At the nominal location, the detector covers the rectangular region $|x|,|y|<12.5~\cm$, corresponding the pseudorapidity range $\eta\gtrsim 8.5$. The FASER$\nu$ detector has the possibility of being moved in the vertical/horizontal directions to correct for changing beam crossing angle orientations.

More recently, the SND@LHC collaboration proposed another neutrino detector to be placed in the tunnel TI18,  which is also $480~\m$ away from the ATLAS interaction IP, but located on its opposite side~\cite{Ahdida:2020evc, Ahdida:2750060}. Notably, the center of the detector would be displaced from the beam collision axis. The detector covers the rectangular region $8~\cm < x < 47~\cm$ and $15.5~\cm < y < 54.5~\cm$, corresponding to the pseudorapidity range $7 \lesssim \eta \lesssim 8.5$. The detector target consists of tungsten and has a mass of 800~kg. 

The baseline beam configuration assumed in this study considers a 13~TeV LHC with a beam half-crossing angle of $150~\murad$ upwards in vertical directions, as used during the end of LHC Run~2. The question on how the fluxes and event rates change for different beam angle configurations will be discussed in \cref{sec:crossing}. When presenting event rates, we assume the nominal integrated luminosity of $150~\ifb$ for Run~3 of the LHC, corresponding to about three years of running. 

\subsection{Event Generation}
\label{sec:event_generator} 

In this study we use and compare the neutrino flux from several commonly used MC event generators. For light hadron production, we use the dedicated cosmic ray and forward physics generators \texttt{EPOSLHC}~\cite{Pierog:2013ria}, \texttt{QGSJET~II-04}~\cite{Ostapchenko:2010vb}, \texttt{DPMJET III.2017.1}~\cite{Roesler:2000he, Fedynitch:2015kcn}, and \texttt{Sibyll~2.3d}~\cite{Ahn:2009wx, Riehn:2015oba, Riehn:2017mfm, Fedynitch:2018cbl}, as implemented in the \texttt{CRMC} simulation package~\cite{CRMC}. In addition, we also simulate light hadrons with the multi-purpose event generator \texttt{Pythia~8.2}~\cite{Sjostrand:2006za, Sjostrand:2014zea} using the \texttt{Monash} tune~\cite{Skands:2014pea}, but note that these predictions have not yet been tuned to or validated with forward physics data. 

To simulate the production of charmed hadrons, we use \texttt{DPMJET III.2017.1}, \texttt{Sibyll~2.3d} and \texttt{Pythia~8.2}. \texttt{Pythia~8.2} provides the option to simulate charm production either as part of their \texttt{SoftQCD} process for minimum bias events, or directly as a \texttt{HardQCD} process, and we will present results for both. We note that only \texttt{Sibyll~2.3d} has been tuned to charm production data. In contrast, neither \texttt{Pythia~8.2} nor \texttt{DPMJET III.2017.1} have been validated for charm production. We mainly include them for comparison, as they have been used in previous studies~\cite{Abreu:2019yak, Ahdida:2020evc}. Indeed, the predictions for forward charm production differ significantly by orders of magnitude between the different generators, as shown in \cref{sec:generator}.

\begin{figure*}[th!]
\includegraphics[width=0.97\textwidth, trim=0 18 0 8, clip]{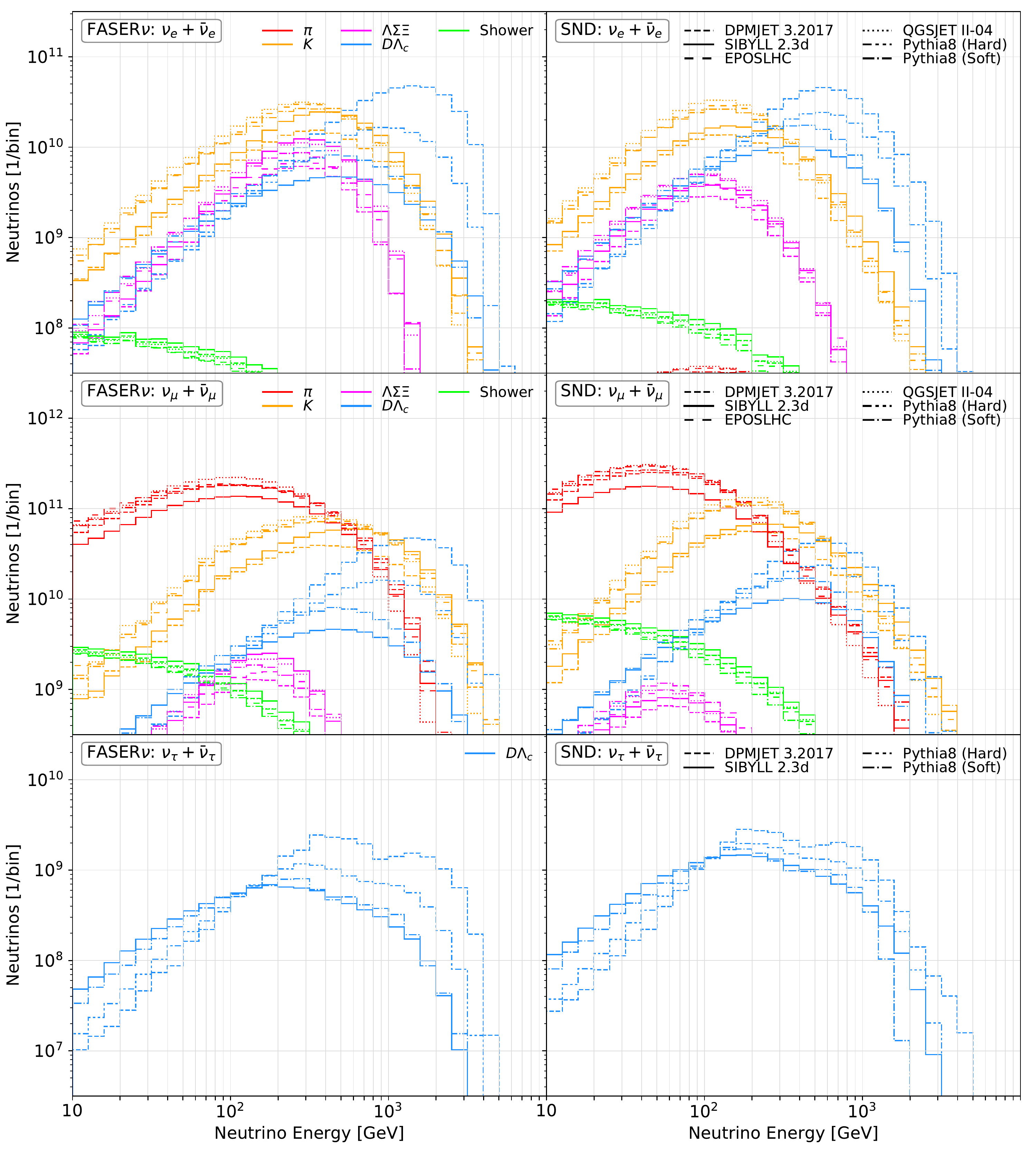}
\caption{\textbf{Neutrino Energy Distribution:} The panels show the neutrino energy spectrum for electron (top), muon (center) and tau (bottom) neutrinos passing through FASER$\nu$ (left) and SND@LHC (right) for LHC Run~3 with an integrated luminosity of $150~\ifb$. The vertical axis shows the number of neutrinos per energy bin that go through the detector's cross sectional area for an integrated luminosity of $150~\ifb$. We separate the different production modes: pion decays (red), kaon decays (orange), hyperon decays (magenta), charm decays (blue) and secondary hadronic showers (green). The different line styles correspond to predictions obtained from \texttt{SIBYLL~2.3d} (solid), \texttt{DPMJET~III.2017.1} (short dashed), \texttt{EPOSLHC} (long dashed), \texttt{QGSJET~II-04} (dotted), and \texttt{Pythia~8.2} using the \texttt{SoftQCD} processes (dot-dashed) and \texttt{Pythia~8.2} with the {HardQCD} process for charm production (double-dot-dashed).}
\label{fig:energy}
\end{figure*}

\subsection{Energy Distribution}
\label{sec:nu_energy} 

\begin{figure*}[th!]
\includegraphics[width=1.0\textwidth]{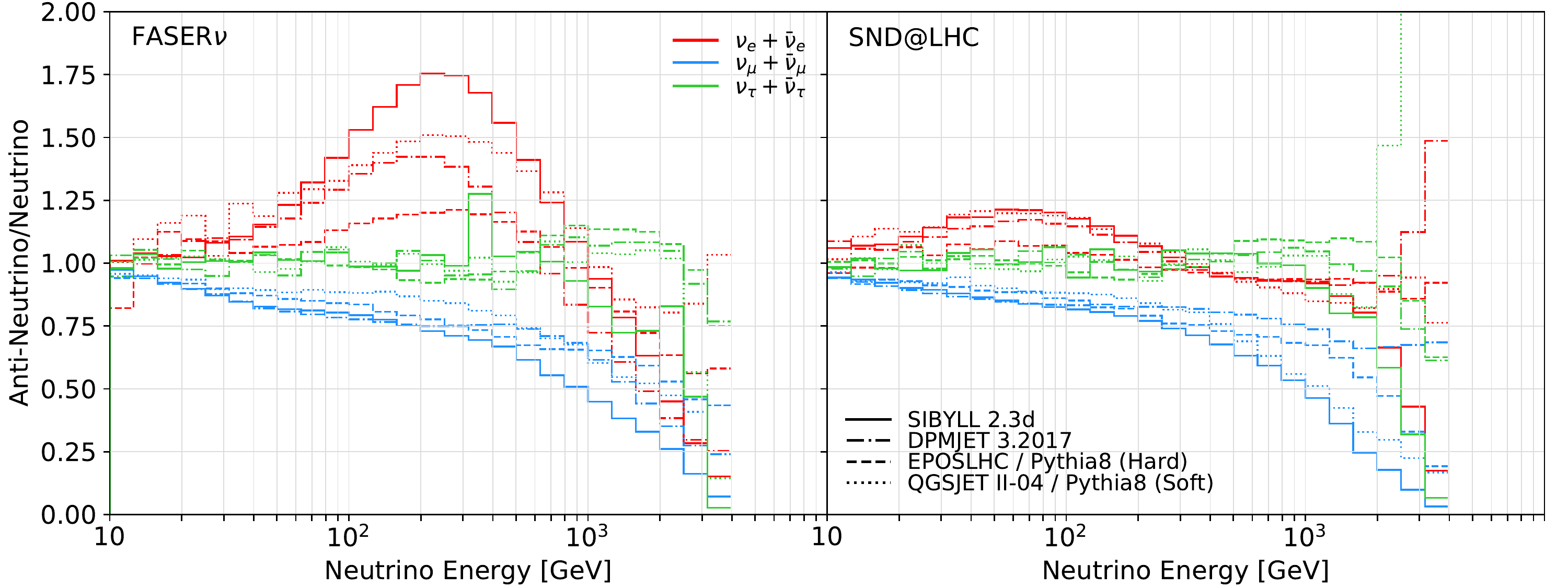}
\caption{\textbf{Neutrino Assymetry:} The panels show the ratio of the anti-neutrino and neutrino fluxes as function of the neutrino energy for electron (red), muon (blue) and tau (green) neutrinos passing through FASER$\nu$ (left) and SND@LHC (right). The different line styles correspond to predictions obtained from \texttt{SIBYLL~2.3d} (solid), \texttt{DPMJET~III.2017.1} (dot-dashed), \texttt{EPOSLHC} for light hadrons / \texttt{Pythia~8.2} using the \texttt{HardQCD} for charm production  (dashed), and \texttt{QGSJET~II-04} for light hadrons / \texttt{Pythia~8.2} using the \texttt{SoftQCD} for charm production (dotted).}
\label{fig:assymetry}
\end{figure*}

Let us now turn to the results of our simulation. In \cref{fig:energy} we show the number of electron (top), muon (center) and tau (bottom) neutrinos going through the cross sectional area of FASER$\nu$ (left column) and SND@LHC (right column) as function of the neutrino energy. 

Our simulation allows us to differentiate the neutrino flux components by origin (in terms of parent particle species), as shown by the differently colored lines. Charged pion decays provide the dominant contribution to the muon neutrino flux at lower energies, while charged kaon decays dominate the muon neutrino flux at higher energies above a few $100~\gev$. Electron neutrinos are mainly produced in kaon decays. While the largest contribution comes from decays of $K_L$ mesons, $K_S$ decays still provide a sizable contribution since their smaller lifetime and hence larger decay-in-flight probability can compensate for the their small semi-leptonic branching fraction. Hyperon decays are generally sub-leading, with the notable exception of anti-electron neutrino production via the decay $\Lambda \to p e^- \bar \nu$. High-energy forward $\Lambda$ production in diffractive scattering is sufficiently enhanced, such that hyperon and kaon decays provide roughly equal contributions at FASER$\nu$. 

Decays of charmed hadrons, including both D-mesons and $\Lambda_c$ baryons, become the dominant production mode for electron and muon neutrinos at the highest energies. Due to the high mass of the tau lepton, tau neutrinos are only produced in the decay of the $D_s$ meson and the subsequent $\tau$ decay.

Neutrinos from the decay of secondary particles produced in downstream hadronic showers only provide a subdominant contribution. These mainly originate from collisions of forward neutral hadrons or diffractive protons with or around the TAN. From a physics point of view, this is a very encouraging and interesting result: it implies that the neutrino flux can be used as an indirect probe of forward hadron production.

\begin{figure*}[th!]
\includegraphics[width=1.0\textwidth]{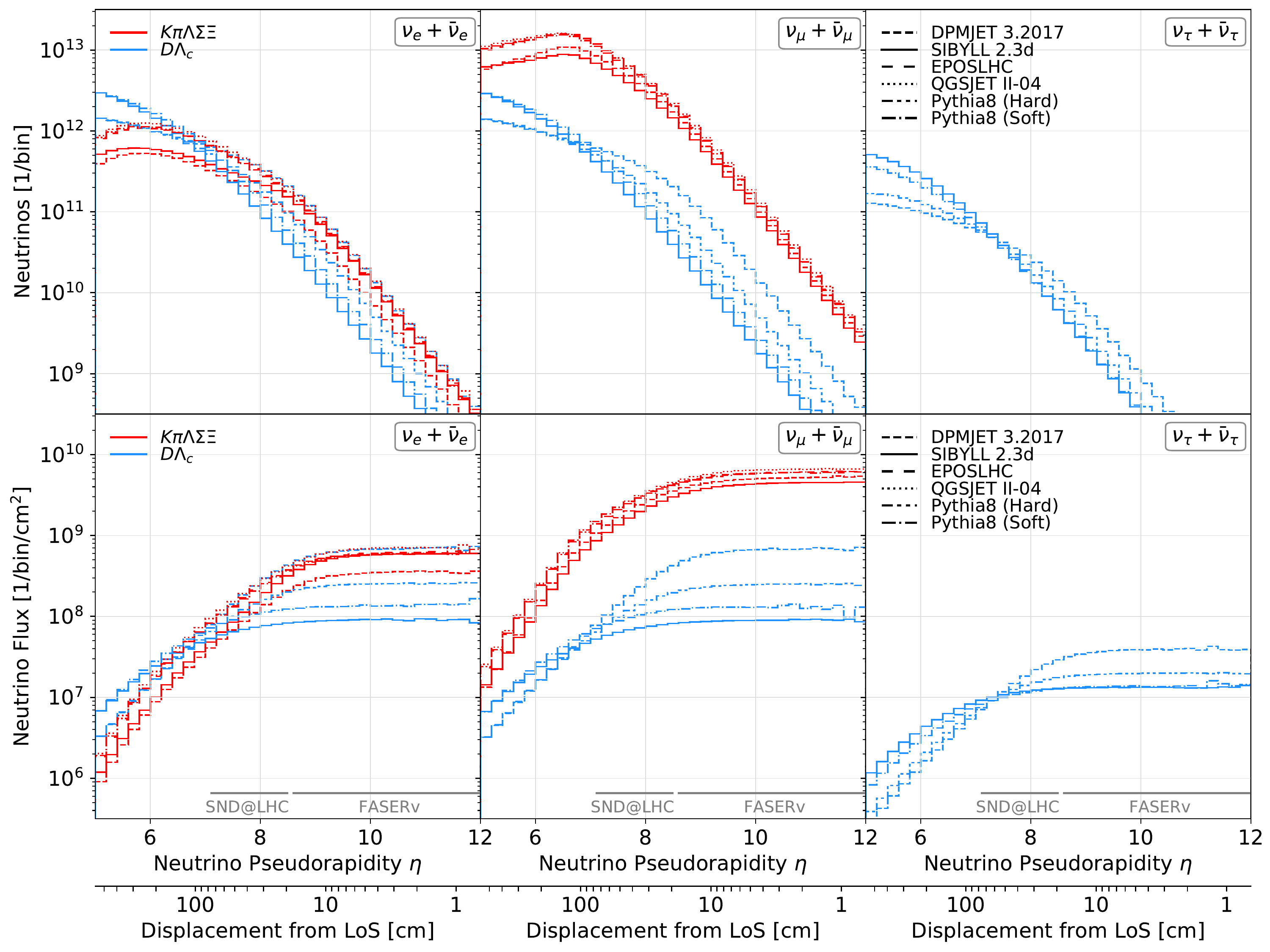}
\caption{\textbf{Neutrino Angular Distribution:} Neutrino pseudorapidity distribution for electron (left), muon (center) and tau (right) neutrinos, as a function of pseudorapidity $\eta$, or equivalently the radial displacement from the line of sight (LoS) at $z = 480~\m$ for LHC Run~3 with an integrated luminosity of $150~\ifb$. In the top panel we show the number of neutrinos, in units of particles per bin, while the bottom panels show the neutrino flux, in units of particles per area per bin. The flux components from light hadron decays and charmed hadron decays are shown in red and blue respectively. The line styles denote the different event generators. All energies $E_\nu > 10~\gev$ are included. Shown at the bottom of each panel is the angular coverage of FASER$\nu$ and SND@LHC.
}
\label{fig:rapidity}
\end{figure*}

In \cref{fig:energy}, we also show the predictions of different generators, as indicated by the different line styles. We can see that there are $\mathcal{O}(1)$ differences between the generator predictions for neutrinos originating from light hadrons. In contrast, there are large differences of more than an order of magnitude on the neutrinos flux from charmed hadron decays. However, as mentioned before, \texttt{DPMJET} and \texttt{Pythia~8} have not been tuned and validated with charm production data, which should be kept in mind when interpreting their predictions. Dedicated efforts are needed, and indeed already ongoing~\cite{Bai:2020ukz}, to provide more reliable predictions for forward charm production. 

\subsection{Flux Asymmetry}
\label{sec:assymetry} 

While not necessarily observable, it is also interesting to look at the differences between neutrino and anti-neutrino production. This is shown in \cref{fig:assymetry} where we plot the ratio of the anti-neutrino and neutrino flux passing through the detectors as function of the neutrino energy. We see that the anti-neutrino and neutrino fluxes are not exactly equal, especially at higher energies. 

This asymmetry is due to small differences between particle and anti-particle production in the far-forward region. Many of the high energy neutrinos come from hadrons that originate from the hadronization of beam remnants, in particular the remaining valence quarks which carry a large fraction of the proton momentum. One therefore expects a larger number of highly energetic hadrons containing an up or down quark than those containing anti-up or anti-down quarks, and hence an asymmetry in the neutrinos produced in their decays.   

A prominent example are charged kaons decays, leading to an enhanced production of neutrinos via the decay $K^+ \to \mu^+ \nu_\mu$ at high energies. Another example is the enhanced forward production of $\Lambda$ baryons compared to $\bar \Lambda$, leading to an enhancement of anti-electron neutrinos via the decay $\Lambda \to p e^- \bar\nu_e$ at intermediate energies $E_\nu \sim \text{few}~100~\gev$. 

In \cref{fig:assymetry}, we also show predictions from different generators. We observe again that there are sizable differences between their predictions. In the case of muon neutrinos, combining measurements of the FASER$\nu$ detector and downstream FASER spectrometer, will allow to measure the muon charge and hence distinguish muon neutrinos and anti-neutrinos and analyze this neutrino flux asymmetry feature.  

\subsection{Angular Distribution}
\label{sec:nu_eta} 

In addition to the energy spectrum, we can also consider the angular distribution of forward LHC neutrinos. This is presented in \cref{fig:rapidity}. The upper panels show the distribution of neutrinos as a function of pseudorapidity for all three flavors. This distribution peaks at intermediate pseudorapidities $\eta \sim 5$ and steeply falls towards higher pseudorapidities as the available phase space steeply decreases. However, the pseudorapidity distribution can be somewhat misleading, since one would need a detector with transverse dimensions that could cover the corresponding area to exploit these neutrinos. If one is interested in maximizing the neutrino event rate for a smaller fixed size detector, a perhaps more interesting quantify is the flux of neutrinos, so the number of neutrinos passing through a unit area. This is shown in the bottom row of plots. The pseudorapidity can also be translated into a displacement from the beam collisions axis (or Line of Sight) at the detector location, which is shown as an additional horizontal scale. At the bottom of each panel, we also illustrate the angular coverage of both detectors.

\renewcommand{\arraystretch}{1.5}
\setlength{\tabcolsep}{5pt}
\begin{table*}[t!]
\centering
\begin{tabular}{c|c||c|c|c||c|c|c}
  \hline
  \hline
  \multicolumn{2}{c||}{Generators} & 
  \multicolumn{3}{c||}{FASER$\nu$} & 
  \multicolumn{3}{c}{SND@LHC} \\
  \hline
  light hadrons & heavy hadrons
  & $\nu_e+\bar\nu_e$    
  & $\nu_\mu+\bar\nu_\mu$    
  & $\nu_\tau+\bar\nu_\tau$    
  & $\nu_e+\bar\nu_e$    
  & $\nu_\mu+\bar\nu_\mu$    
  & $\nu_\tau+\bar\nu_\tau$    
  \\
  \hline
    SIBYLL & SIBYLL
  & 901   
  & 4783  
  & 14.7  
  & 134   
  & 790   
  & 7.6   
  \\
    DPMJET & DPMJET
  & 3457  
  & 7088  
  & 97    
  & 395   
  & 1034  
  & 18.6  
  \\
    EPOSLHC & Pythia8 (Hard)    
  & 1513  
  & 5905  
  & 34.2  
  & 267   
  & 1123  
  & 11.5  
  \\
    QGSJET & Pythia8 (Soft)    
  & 970   
  & 5351  
  & 16.1  
  & 185   
  & 1015  
  & 7.2   
  \\
  \hline 
    \multicolumn{2}{c||}{Combination (all)}
  & $1710^{+1746}_{-809}$  
  & $5782^{+1306}_{-998}$  
  & $40.5^{+56.6}_{-25.8}$  
  & $245^{+149}_{-111}$  
  & $991^{+132}_{-200}$  
  & $11.3^{+7.3}_{-4.0}$  
  \\
  \hline
    \multicolumn{2}{c||}{Combination (w/o DPMJET)}
  & $1128^{+385}_{-227}$  
  & $5346^{+558}_{-563}$  
  & $21.6^{+12.5}_{-6.9}$  
  & $195^{+71}_{-61}$  
  & $976^{+146}_{-185}$  
  & $8.8^{+2.7}_{-1.5}$  
  \\
  \hline
  \hline
\end{tabular}
\caption{Expected number of charged current neutrino interaction events occurring in FASER$\nu$ and SND@LHC during LHC Run~3 with $150~\ifb$ integrated luminosity. Here we assume a target mass of $1.2$ tons for FASER$\nu$ and $800~\kg$ for SND@LHC; further details on the experimental setup are provided in \cref{sec:experiment}. We provide predictions for \texttt{SIBYLL~2.3d}, \texttt{DPMJET~III.2017.1}, \texttt{EPOSLHC}/\texttt{Pythia~8.2} with \texttt{HardQCD}, and \texttt{QGSJET~II-04}/\texttt{Pythia~8.2} with \texttt{SoftQCD}. The two bottom rows provide a combined average, both including and excluding the \texttt{DPMJET} prediction, where the uncertainties correspond to the range of predictions obtained from different MC generators. }
\label{tab:interactions}
\end{table*}

\begin{figure*}[th!]
\includegraphics[width=1.0\textwidth]{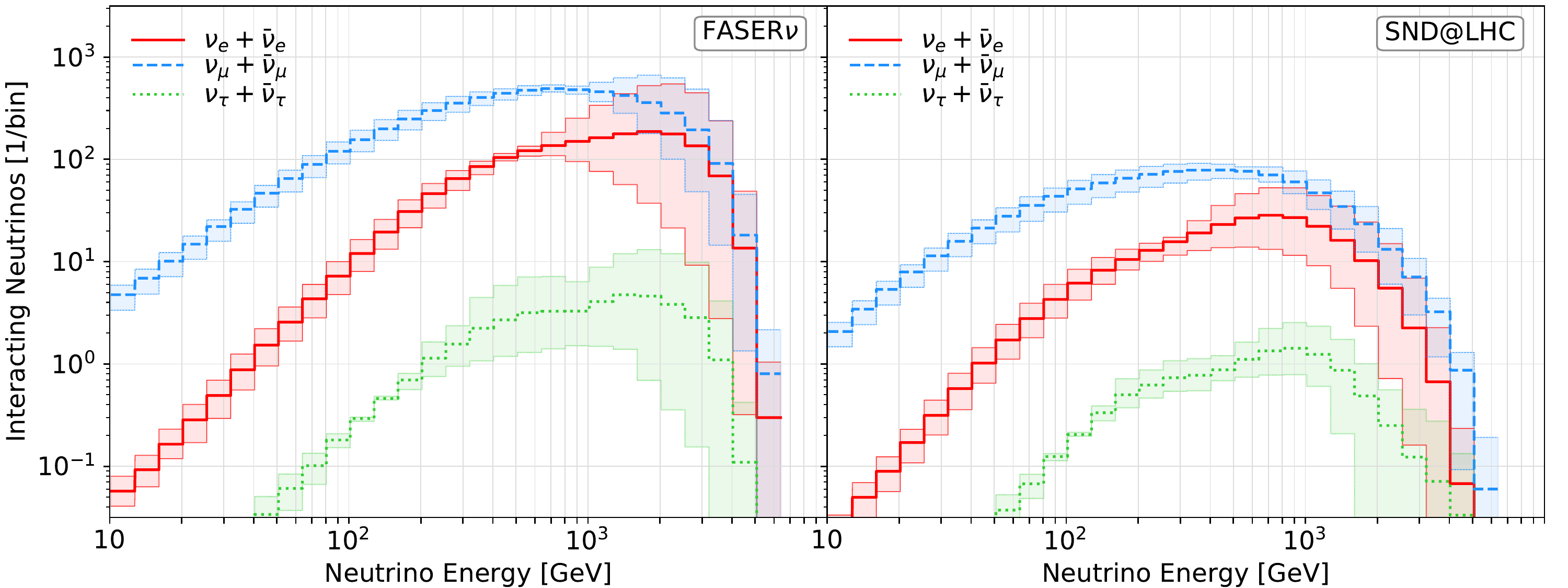}
\caption{\textbf{Interacting Neutrino Energy Distribution:} The panels show the number of charged current neutrino interactions with the FASER$\nu$ (left) and SND@LHC (right) detectors during LHC Run~3 with $150~\ifb$ integrated luminosity as function of the neutrino energy. Here we assume a target mass of $1.2$ tons for FASER$\nu$ and $800~\kg$ for SND@LHC; further details on the experimental setup are provided in \cref{sec:experiment}. The red solid, blue dashed and green dotted lines correspond to electron, muon and tau neutrinos, respectively. The thick line denotes to the average prediction of different generators, while the shaded band corresponds to the range of predictions obtained with different generators.}
\label{fig:interaction}
\end{figure*}

For all neutrino flavors, the neutrino flux peaks around the beam collision axis and falls off when moving away from it. This is both due to the angular spectrum of the parent hadrons as well as due to the LHC's long but narrow beam pipe shape. The number of neutrino events per detector mass can therefore be maximized for an experiment placed right on the beam collision axis, which is the case for FASER$\nu$. Here the neutrino flux is almost constant throughout the detector's cross sectional area. In contrast, SND@LHC is displaced from the beam collision axis and the neutrino flux drops considerably between from the high and low pseudrorapidity ends of the detector, which indicates the potential to probe the pseudrorapidity dependence of the neutrino flux in this region. 

As before, the different colors correspond to the different production mechanisms: light hadron decays in red, charmed hadron decays in blue and downstream hadronic showers in green. The line styles denote the different event generators. Similar to what we have seen for the energy spectrum, we observe $\mathcal{O}(1)$ differences between the MC generator predictions for neutrinos from light hadron decays, while the differences for neutrinos from charm decay are significantly larger. 

\section{Neutrino Event Rates}
\label{sec:interactions} 

\subsection{Interactions}
\label{sec:inter} 

While we have so far concentrated on the number of neutrinos passing though the detector, let us now discuss the number of neutrinos interacting with each detector. For this, we convolute the previously obtained neutrino flux with the LO charged current neutrino interaction cross section with a tungsten target obtained via GENIE~\cite{Andreopoulos:2009rq, Andreopoulos:2015wxa} where we have included a suppression factor for the tau neutrino cross section obtained in Ref.~\cite{Kretzer:2002fr}. 

The resulting number of expected charged current neutrino interactions in FASER$\nu$ and SND@LHC during LHC Run~3 with an integrated luminosity are presented in \cref{tab:interactions}. Since not all generators are able to both simulate light hadron and charm production, we group them together in four setups: i) \texttt{SIBYLL~2.3d}, ii) \texttt{DPMJET~III.2017.1}, iii)  \texttt{EPOSLHC} for light hadrons and \texttt{Pythia~8.2} with \texttt{HardQCD} for charmed hadrons, and iv)  \texttt{QGSJET~II-04} for light hadrons and \texttt{Pythia~8.2} with \texttt{SoftQCD} for charmed hadrons. As before, we observe sizable differences between the different MC generator prediction, which are mainly related to the neutrino flux from charmed hadron decays. The lowest and highest event rates are predicted by \texttt{SIBYLL~2.3d} and \texttt{DPMJET~III.2017.1}, respectively. Notably, the predictions for the tau neutrino event rate in FASER$\nu$ differ by more than a factor six, ranging from 21 (\texttt{SIBYLL}) to 131 (\texttt{DPMJET}). 

In the two bottom rows, we combine these four setups to an average prediction, where the uncertainty corresponds to the range of predictions provided by the different generators. Since \texttt{DPMJET} provides significantly larger charm production predictions than the other generators, we also provide a separate average excluding the \texttt{DPMJET} prediction in the last row. We want to emphasize again that neither \texttt{DPMJET} nor \texttt{Pythia~8.2} have been validated or tuned with charm data, and their predictions should therefore be taken with a grain of salt. Better estimates for forward charm production are needed. In addition, the range of generator predictions are only a crude measure for the neutrino flux uncertainties and should ideally be replaced by individual uncertainty prediction provided by the generators. 
\renewcommand{\arraystretch}{1.5}
\setlength{\tabcolsep}{5pt}
\begin{table*}[t!]
\centering
\begin{tabular}{c||c|c|c||c|c|c||c|c|c}
  \hline
  \hline
  Beam Configuration &
  \multicolumn{3}{c||}{FASER$\nu$} & 
  \multicolumn{3}{c||}{FASER$\nu$ (adjusted)} & 
  \multicolumn{3}{c}{SND@LHC} \\
  \hline
  Crossing Angle
  & $\nu_e+\bar\nu_e$    
  & $\nu_\mu+\bar\nu_\mu$    
  & $\nu_\tau+\bar\nu_\tau$    
  & $\nu_e+\bar\nu_e$    
  & $\nu_\mu+\bar\nu_\mu$    
  & $\nu_\tau+\bar\nu_\tau$   
  & $\nu_e+\bar\nu_e$    
  & $\nu_\mu+\bar\nu_\mu$    
  & $\nu_\tau+\bar\nu_\tau$   
  \\
  \hline
    +150$\mu$rad vertical
  & 901  
  & 4783  
  & 14.7 
  & 996  
  & 5154  
  & 13.0  
  & 134  
  & 790  
  & 7.6  
  \\
    -150$\mu$rad vertical
  & 912  
  & 4802  
  & 16.3 
  & 965 
  & 5141 
  & 15.2  
  & 100  
  & 520  
  & 6.3  
  \\
    +150$\mu$rad horizontal   
  & 953  
  & 5095 
  & 19.7 
  & 1027  
  & 5221  
  & 16.3 
  & 129  
  & 705  
  & 7.4  
  \\
   -150$\mu$rad horizontal    
  & 921  
  & 4912  
  & 16.6 
  & 986  
  & 5167  
  & 13.2 
  & 109  
  & 586  
  & 6.9  
  \\
  \hline
   no crossing angle
  & 989  
  & 5389  
  & 15.3 
  & --- 
  & --- 
  & --- 
  & 118  
  & 646  
  & 6.9  
  \\
  \hline
  \hline
\end{tabular}
\caption{Expected number of charged current neutrino interactions in FASER$\nu$ and SND@LHC during LHC Run~3 with $150~\ifb$ integrated luminosity for different orientations of the beam crossing angle. The simulations were performed using \texttt{Sibyll~2.3d}. For FASER$\nu$, we provide both predictions for its nominal position centered around $x=y=0$ and an adjusted position moved $7.2~\cm$ into the same direction as the beam.}
\label{tab:xing}
\end{table*}

In \cref{fig:interaction}, we present the energy spectrum of neutrinos interacting with the detectors. Compared to the energy spectrum presented in \cref{fig:energy}, this spectrum is shifted towards higher energy, since the neutrino interaction cross section roughly increases linearly with the neutrino beam energy.  The shaded band corresponds to the range of predictions obtained from the different MC event generators, while the thick central line shows their average. As expected, the uncertainties are largest at higher energies and for tau neutrinos where the charm production mode dominates. 

We note that there are a variety of other uncertainties on the expected number of neutrino interactions, which are not included in this study. As already mention before, this includes the modeling of the secondary neutrino flux component. While it was found to be sub-leading for higher energies neutrinos in the far-forward direction, it is expected to become more important at lower energies. In addition, there are also uncertainties associated with neutrino interactions, for example associated with nuclear PDFs (including showing, anti-shadowing and the EMC effect), the hard scattering cross section, and final state hadronic effect (including parton shower, hadronization, response of the nuclear medium on the developing shower, final state interactions). Dedicated studies are needed, and in part already ongoing~\cite{Anchordoqui:2021ghd}, to understand, quantify and reduce the associated uncertainties.

\begin{figure*}[th!]
\includegraphics[width=1.0\textwidth]{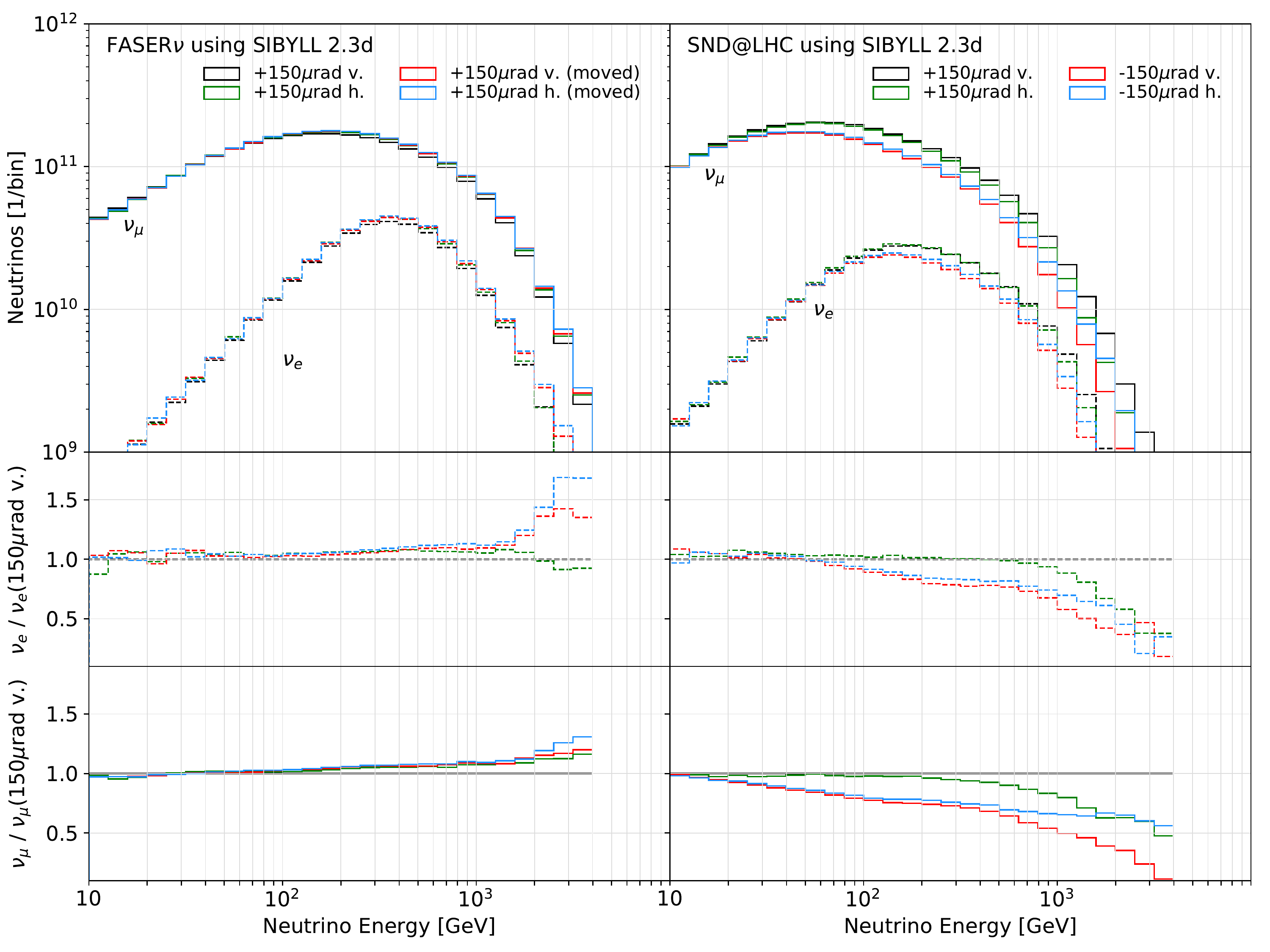}
\caption{\textbf{Beam Configuration}: We show the neutrino flux going through FASER$\nu$ (left) and SND@LHC (right) for different orientations of the beam crossing angle considered for LHC Run~3 with an integrated luminosity of $150~\ifb$. The simulations were performed using \texttt{Sibyll~2.3d}. The muon neutrino and electron neutrino fluxes are shown as solid and dashed lines, respectively. The lower panels show the change of the neutrino flux relative to the nominal setup with an angle of $150~\murad$ vertically upwards as shown in black. For FASER$\nu$, we also show results for a detector that was moved to be aligned with the center of the neutrino beam. 
}
\label{fig:xing}
\end{figure*}

\subsection{Crossing Angles}
\label{sec:crossing} 

To avoid parasitic collisions in the beam pipe away from the IP, the LHC's proton beams have a small beam crossing angle when they collide. So far, we have assumed a beam configuration at the ATLAS IP similar to the end of LHC Run~2 with a beam half-crossing angle of about $150~\murad$ vertically upwards. To distribute the collision debris more evenly and allow for a longer lifetime of the LHC's focusing magnets, changing orientations of the beam crossing angle are considered for Run~3 of the LHC.  

At the detector locations $480~\m$ downstream from the IP, a beam crossing angle of $150~\murad$ leads to a shift of the beam collision axis of $7.2~\cm$ relative to its nominal position assuming no crossing angle. Different beam crossing angle orientations can therefore shift the neutrino beam and change the expected neutrino event rate. To analyze this effect, we run our simulation for four different beam orientations (upward, downward, left and right) with a $150~\murad$ half-crossing angle. For comparison, we also consider the scenario with perfectly parallel beams and no crossing angle. 

The resulting number of neutrino interactions with the detectors are shown in \cref{tab:xing}, where we have used \texttt{Sibyll~2.3d} as event generator. We can see that the event rates at FASER$\nu$, which is centered around the nominal beam collision axis at $x=y=0$, are only marginally effected as expected from its symmetric setup. In contrast, the event rates at SND@LHC, which is located in the $x,y>0$ quadrant, sensitively depend on the beam orientation. In particular, when changing from a vertically upward to a downward orientation, the event rates drop by about $30~\%$. 

The FASER$\nu$ detector also has the option of being moved around its nominal position, such that it can be aligned with the center of the neutrino beam. As also shown in \cref{tab:xing}, such an alignment would slightly increase the expected event rates.  

In \cref{fig:xing}, we show the energy spectra of electron neutrinos (dashed lines) and muon neutrinos (solid lines) passing through the detectors for different choices of the crossing angles. The lower panels show the change with respect to the baseline orientation vertically upwards. For FASER$\nu$, we show results for beam orientations $+150~\murad$ vertically and horizontally for the nominal detector location centered around $x=y=0$, and the adjusted location centered around $x$ or $y=7.2~\cm$. A mildly increased flux of neutrino high energy neutrinos with $E\gtrsim 1~\tev$ is observed for the adjusted detector location compared to the nominal one. For SND@LHC, we consider all four orientations of beam crossing angle. A sizable reduction of the neutrino flux at high energies compared to the upward beam orientation is observed for horizontal or downward setups. However, while a change of the beam orientation reduced the overall event rate, it also allows SND@LHC to probe the neutrino spectrum in a larger effective pseudorapidity range. 

\section{Summary}
\label{sec:summary} 

Starting in 2022, two dedicated experiments to detect neutrinos produced in LHC collisions, FASER$\nu$ and SND@LHC, will start to take data in the far-forward region of the LHC. This emerging LHC neutrino physics program requires reliable estimates of the LHC’s forward neutrino fluxes and their uncertainties. Such estimates currently rely on sophisticated but rather computationally expensive and time consuming computation performed by dedicated groups using BDSIM or FLUKA. In this work, we presented an alternative fast neutrino flux simulation, implemented as a RIVET module, which is accessible to the entire community and can be run on a personal computer within minutes.

We have used this simulation to obtain the neutrino flux going through both neutrino detectors for all flavors using six commonly used event generators. We have presented the energy and pseudorapidity spectrum for the neutrino flux, and found sizable difference between the generator predictions, especially for neutrinos from charmed hadron decays. We note again that  dedicated efforts are needed, and have already started~\cite{Bai:2020ukz}, to provide more reliable predictions for this production mode. 

This module is also well suited for phenomenological studies and applications. As an example, we have studied the impact of different beam crossing angle orientations as well as detector locations for FASER$\nu$. Since the presented neutrino flux simulation is already implemented as a RIVET module, it would also be ideally suited for generator tuning applications in the future when data becomes available. As we have seen, the neutrino flux for different flavors, at different energies and in different rapidity regions (or experiments) is sensitive to different flux components and hence provides information on the forward production for various parent hadron species. Given the high expected event rates, this would be valuable data to constrain forward particle production, which plays an important role in astroparticle physics~\cite{Engel:2011zzb, SnowmassCosmicRay}. For example, this data could help to understand the observed excess of muons in cosmic-ray air showers~\cite{Dembinski:2019uta, SnowmassMuonExcess} and to constrain the prompt atmospheric neutrino background at large-scale neutrino telescopes~\cite{SnowmassCosmicNeutrino}. 

\section*{Acknowledgements}

We thank Akitaka Ariga, Tomoko Ariga, Jamie Boyd, Tomohiro Inada, Anatoli Fedynitch, Jonathan Feng, Max Fieg, Tanguy Pierog, Holger Schulz, and Sebastian Trojanowski for useful discussions.  We are  particularly grateful to Francesco Cerutti and Marta Sabaté Gilarte from the CERN Sources, Targets and Interactions Group for sharing the results of their dedicated FLUKA simulation of the LHC neutrino flux. We are also grateful to the authors and maintainers of many open-source software packages, including
\texttt{CRMC}~\cite{CRMC},
\texttt{RIVET}~\cite{Buckley:2010ar},
\texttt{scikit-hep}~\cite{Rodrigues:2019nct}, and
\texttt{uproot}~\cite{jim_pivarski_2019_3256257}.
F.K. is supported by the U.S. Department of Energy under Grant No. DE-AC02-76SF00515. The development of LHC models using BDSIM by L.N. has been supported by the Science and Technology Research council grant ``The John Adams Institute for Accelerator Science'' ST/P00203X/1 and ST/V001620/1 through the John Adams Institute at Royal Holloway.

\appendix

\begin{figure*}[th!]
\includegraphics[width=1.0\textwidth]{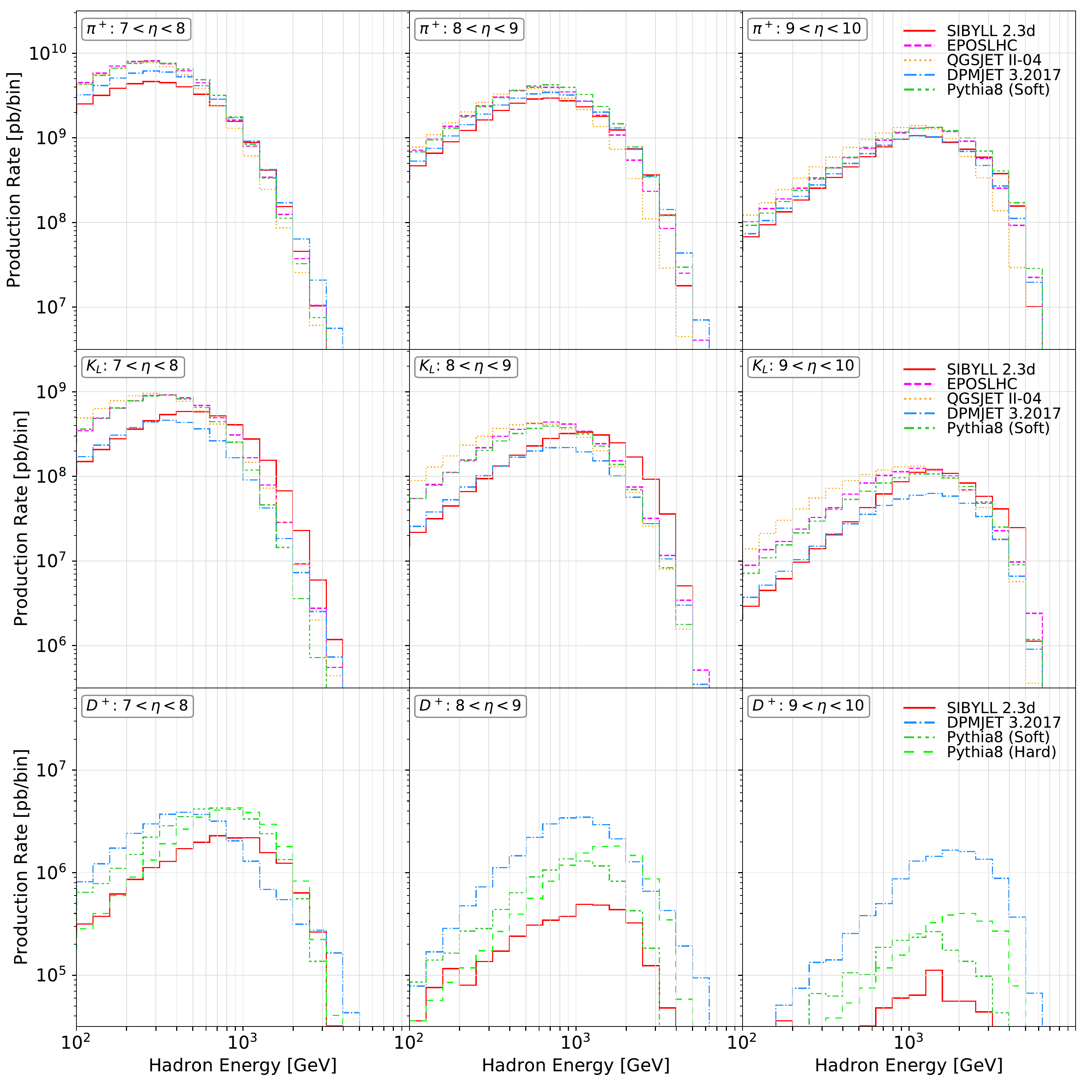}
\caption{Predictions for the energy spectra of $\pi^+$ (top row), $K_L$ (center row) and $D^+$ (bottom row) in three pseudorapidity intervals using six commonly used MC generators for LHC Run~3 with an integrated luminosity of $150~\ifb$.}
\label{fig:generators}
\end{figure*}

\section{Event Generator Comparison for Forward Particle Production}
\label{sec:generator} 

In the main part of this paper, we have compared the neutrino flux predictions provided by different MC event generators. In this appendix, we also compare generator predictions for  forward hadrons production rates directly.  

This is done in \cref{fig:generators} for three most important hadrons, $\pi^+$ (top), $K_L$ (center) and $D^+$ (bottom), in three different far-forward pseudorapidity bins. These results are consistent with our previous findings. The generator predictions are consistent up to an $\mathcal{O}(1)$ factor for light hadron production, although sizable shape difference can be seen. In contrast, there are large differences of an order of magnitude and more for charm production at high pseudorapidities $\eta>8$. 

\section{Comparison to FLUKA}
\label{sec:fluka} 

\begin{figure}[th!]
\includegraphics[width=0.45\textwidth]{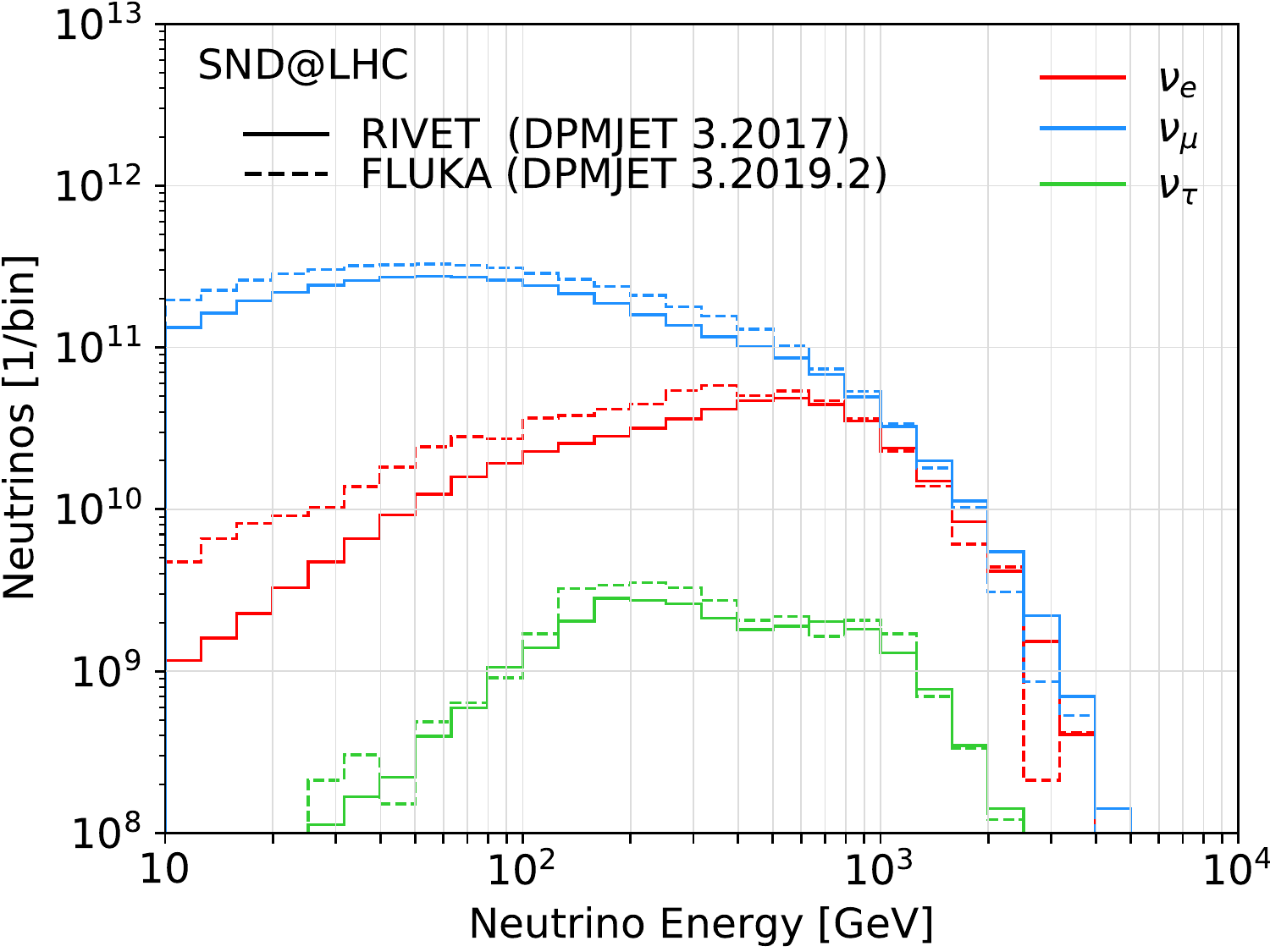}
\caption{Comparison of the neutrino flux obtained using the far neutrino flux simulation implemented as RIVET module (solid) and the a full FLUKA simulation (dashed) using \texttt{DPMJET~III} generator for LHC Run~3 with an integrated luminosity of $150~\ifb$. The FLUKA simulation was performed by CERN Sources, Targets and Interactions Group in the context of the XSEN proposal~\cite{XSEN:2019bel}, and its results have further been presented in Refs.~\cite{Beni:2020yfy, Ahdida:2020evc, Ahdida:2750060}. See text for discussion. 
}
\label{fig:fluka}
\end{figure}

In \cref{sec:fnfs-validation} we have compared the result of the fast neutrino flux simulation against a full simulation performed with BDSIM, and found good agreement of both predictions. 

In addition, we have also compared our simulation against another full simulation using FLUKA. This simulation was performed by the CERN Sources, Targets and Interactions Group in the context of the XSEN proposal~\cite{XSEN:2019bel, Beni:2020yfy} and was also used for the SND@LHC proposal~\cite{Ahdida:2020evc, Ahdida:2750060}. Similar to the setup used throughout this study, the FLUKA simulation assumes the end of LHC Run 2 configuration with a center of mass energy of $13~\tev$ and a beam half-crossing angle of $150~\murad$ vertically upwards. Following the envisioned geometrical coverage of XSEN, the simulation results recorded neutrinos passing through a cross sectional area with $|x| \!<\! 70~\cm$ and $5~\cm \!<\! y \!<\! 70~\cm$. While this area does not enclose FASER$\nu$. 

This is shown in \cref{fig:fluka}, where we plot the energy spectrum of neutrinos passing through the cross sectional area of SND, defined by $8~\cm \!<\! x \!<\! 47~\cm$ and $15.5~\cm \!<\! y \!<\! 54.4~\cm$. The results obtained using the fast simulation implemented as RIVET module and the full FLUKA simulation are shown as solid and dashed histograms, respectively. We see generally a good agreement at higher energies, but note some differences at lower energies. Similar to our comparison with BDSIM, this could indicate the existence of further contributions to the secondary neutrino flux component which are not included in the fast simulation. 

However, let us also note some differences between the full and fast simulation that could contribute to the differences: i) Different versions of the event generator were used. While the fast simulation uses \texttt{DPMJET~3.2017} as implemented in \texttt{CRMC}, the FLUKA simulation uses the newer version \texttt{DPMJET~3.2019.2} for which internal parameters associated with fragmentation have been changed~\cite{DPMJET}. ii) The decay of particles is modeled using \texttt{Pythia} in the fast simulation and inside FLUKA for the full simulation. iii) In the fast simulation all primary interactions occur at the nominal interaction point, while in the FLUKA simulation the collisions are sampled around the nominal interaction point with a longitudinal spread following a Gaussian distribution with a width of a few cm. These and other modeling aspects should be aligned to allow for more fair comparisons in the future. 


\bibliography{references}

\end{document}